# Dependence of microwave absorption properties on ferrite volume fraction in MnZn ferrite/rubber radar absorbing materials

Adriana M. Gama, Mirabel C. Rezende and Christine C. Dantas

*Divisão de Materiais (AMR), Instituto de Aeronáutica e Espaço (IAE),*
*Departamento de Ciência e Tecnologia Aeroespacial (DCTA), Brazil*

Abstract

We report the analysis of measurements of the complex magnetic permeability ($\mu_r$) and dielectric permittivity ($\varepsilon_r$) spectra of a rubber radar absorbing material (RAM) with various MnZn ferrite volume fractions. The transmission/reflection measurements were carried out in a vector network analyzer. Optimum conditions for the maximum microwave absorption were determined by substituting the complex permeability and permittivity in the impedance matching equation. Both the MnZn ferrite content and the RAM thickness effects on the microwave absorption properties, in the frequency range of 2 to 18 GHz, were evaluated. The results show that the complex permeability and permittivity spectra of the RAM increase directly with the ferrite volume fraction. Reflection loss calculations by the impedance matching degree (reflection coefficient) show the dependence of this parameter on both thickness and composition of RAM.

*Keywords:*

*Email address*: adrianaamg@iae.cta.br; mirabelmcr@iae.cta.br; christineccd@iae.cta.br;
(Adriana M. Gama, Mirabel C. Rezende and Christine C. Dantas )

# 1. Introduction

It is well known that the development of radar absorbing materials (RAM) is fundamental in stealth technology, as well as in other applications in the microwave range, where the reduction of electromagnetic interference and the solution of electromagnetic compatibility problems are necessary. In order to design an adequate RAM, the complex permeability ($\mu_r = \mu' - i\mu''$) and permittivity ($\varepsilon_r = \varepsilon' - i\varepsilon''$) are fundamental physical quantities for determining the reflection or attenuation properties of the material.

In the case of ferrite-polymer composites, several studies have been carried out to investigate the effects of ferrite materials and their volume fractions on the microwave absorbing properties [1, 2]. In a similar way, the influence of conducting materials addition has been investigated [3]. Some attempts were also made to verify the correlation between the material constants ($\mu_r$ and $\varepsilon_r$) and the microwave absorption in the UHF/VHF frequencies [4, 5].

Composite materials are useful as microwave absorbers due to their advantages in respect to lighter weight, lower cost, design flexibility, and adjustable microwave properties over intrinsic ferrites. Usually, ferrite and ferrite composites backed with a conducting plate are used to achieve absorption [4, 6, 7].

In the case of a metal-single layered absorber, the normalized input impedance with respect to the impedance in free space, $Z$, and reflection loss ($RL$) with respect to the normal incident plane wave in a rectangular waveguide are given by [4, 7]:

$$Z = \sqrt{\mu_r/\varepsilon_r} \tanh(-i\frac{2\pi}{\lambda}t\sqrt{\mu_r\varepsilon_r}) \qquad (1)$$

$$RL \text{ (dB)} = 20 \log_{10}\left|\frac{Z-1}{Z+1}\right| \qquad (2)$$

where: $\lambda$ is the wavelength of the incident plane wave in free space, and $t$ is the sample thickness. The impedance matching condition representing the perfect absorbing properties is given by $Z \to 1$. The impedance matching condition is determined by combination of six parameters $\mu'$, $\mu''$, $\varepsilon'$, $\varepsilon''$, $\lambda$ and $t$. Their relationships for zero reflection is not simple, and numerical techniques can provide support to simulate the matching conditions.

In the present paper, the complex magnetic permeability and dielectric permittivity for rubber radar absorbing materials with various MnZn ferrite volume fractions are reported. The effects of different MnZn ferrite content and the thickness values of RAM on the microwave absorption properties in the frequency range of 2 - 18 GHz are also discussed.

# 2. Experimental

The composite samples evaluated in the present investigation were prepared by using a MnZn ferrite powder supplied by Sontag S.A. Company from Brazil and a silicone rubber from BASF Ltd. as a polymeric matrix. The MnZn ferrite particles were prepared by usual ceramic sintering method from the mixture of $Fe_2O_3$, $MnCO_3$ and $ZnO$. The stoichiometry of the prepared MnZn ferrite is $Mn_{0.66}Zn_{0.34}Fe_2O_4$, which has the spinel structure with a lattice constant $a = 8.483$ Å.

The densities of the MnZn ferrite and the silicon matrix of the present work are 5.029 and 1.28 g/cm$^3$, respectively. The volume fraction of the ferrite in the silicon matrix varied in the range $v_f = 0, 0.10, 0.15, 0.20, 0.27$ and $0.37$.

Elastomeric specimens filled with the magnetic powder were prepared by conventional mechanical mixture of the raw materials. The homogeneous mixtures were molded in a coaxial die with inner diameter of 3 mm, an outer diameter of 7 mm and a length of 5 mm. The polymer curing was performed at room temperature for about 24 hours. At the end, flexible cylindrical composite specimens were produced.

Scanning electron microscopy (SEM) examinations were performed employing a DSM 950 Zeiss, without special preparation of the samples, and X ray diffraction (XRD) spectra of powders were obtained using Cu K$\alpha$ radiation from an PW 1830 Philips X-ray diffractometer, and the diffraction points were recorded from $10^0$ to $80^0$.

The scattering parameters (S parameters) were measured and used to calculate the complex magnetic permeability and dielectric permittivity of the prepared RAM [8 - 11]. The measurements were performed according to the transmission/reflection method using an HP 8510C vector network analyzer, adapted with an APC7 coaxial transmission line [10], in the frequency range of 2–18 GHz.

## 3. Results and discussion

Figs. 1 and 2 show the SEM image and the XRD patterns of the MnZn ferrite, respectively. Fig. 1 presents the varieties of particle sizes (1 – 50 µm) and shapes (acicular or plate) of the MnZn ferrite.

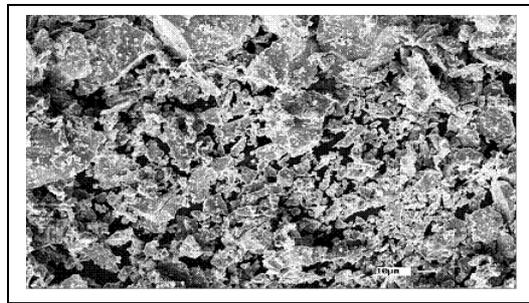

Fig.1. A SEM photography of the MnZn ferrite used.

The X-ray diffraction pattern for the system $Mn_{0,66}Zn_{0,34}Fe_2O_4$ powder (Fig. 2) shows the existence of a single cubic phase. The contributions related to the crystal structure were found to be in good agreement with those obtained by a JCPDS card (74-2401) for the MnZn ferrite.

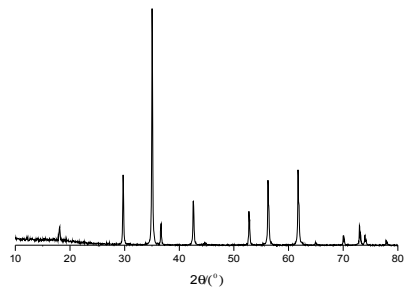

Fig.2. XRD pattern of the $Mn_{0,66}Zn_{0,34}Fe_2O_4$ ferrite.

Figs. 3 and 4 show the measured values of the real and imaginary permeability ($\mu$' and $\mu$'') and permittivity ($\varepsilon$' and $\varepsilon$'') quantities, respectively, as a function of the frequency, for the MnZn ferrite-rubber composites at ferrite volume fractions of $v_f$ = 0, 0.10, 0.15, 0.20, 0.27 and 0.37. The real components of $\mu$ and $\varepsilon$ parameters of the pure silicone rubber ($v_f$ = 0) are nearly constant in the evaluated frequency range, with values of 1 and 3, respectively. The imaginary components of $\mu$ and $\varepsilon$ present lower values, that vary from 0 to 1.2 and 0 to 0.7, respectively. These behaviors mean that the pure silicon rubber presents low magnetic and dielectric losses [3, 4].

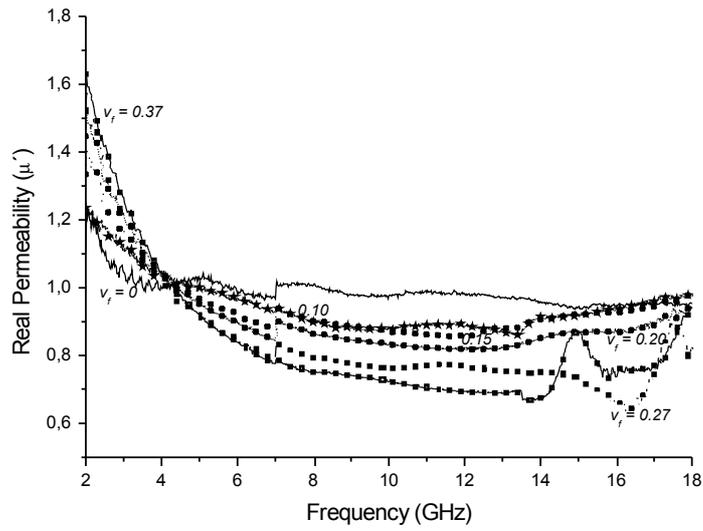

**(a)**

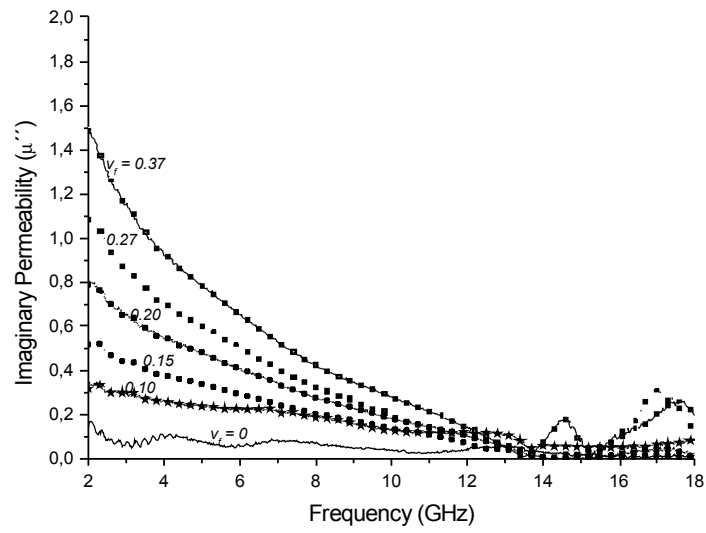

**(b)**

Fig. 3. Real (a) and imaginary (b) permeability for MnZn ferrite-rubber composites at $v_f$ = 0, 0.10, 0.15, 0.20, 0.27 and 0.37.

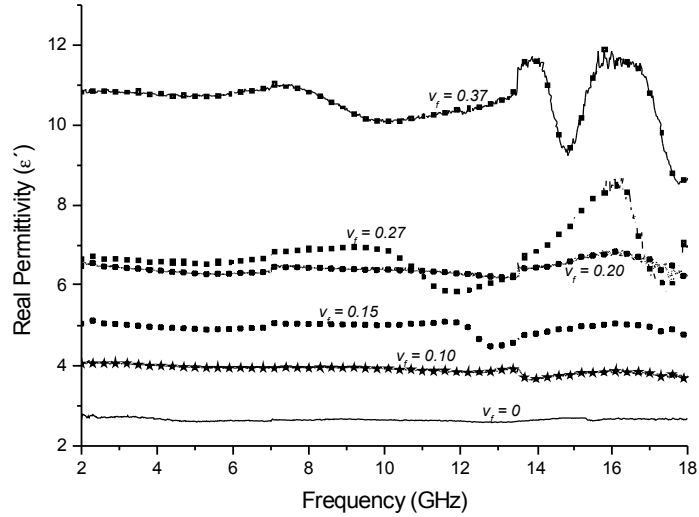

(a)

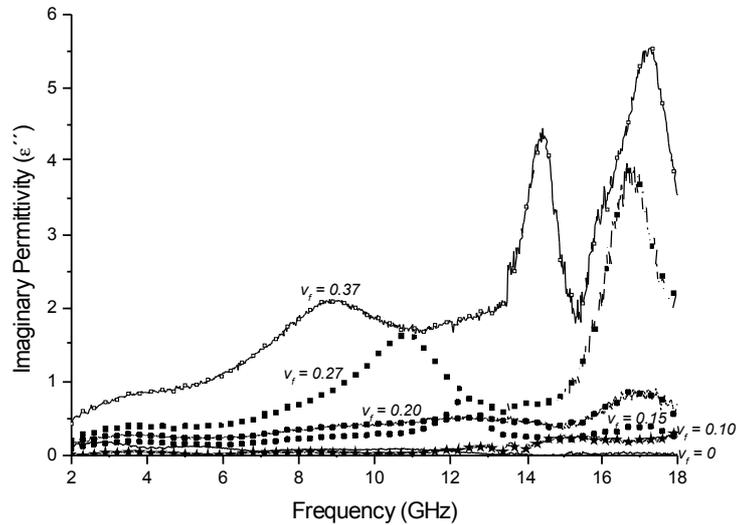

(b)

Fig. 4. Real (a) and imaginary (b) permittivity for MnZn ferrite-rubber composites at $v_f$ = 0, 0.10, 0.15, 0.20, 0.27 and 0.37.

The complex permittivity spectra of the composites generally reach increasingly higher values as the ferrite volume fraction increases, with a few exceptions at higher frequency ranges (above 14 GHz, see Fig. 4). This trend is also followed by the complex permeability spectra, but only for lower frequencies ($\leq$ 5 GHz) in the case of $\mu'$ (Fig. 3(a)) and up to ~12 GHz for $\mu''$ (Fig. 3(b)). The $\mu'$ and $\mu''$ values decrease with the frequency increase, with an exponential-like behavior. The $\varepsilon'$ is nearly independent of the frequency, however, the $\varepsilon''$ varies in a very complicated manner. In both, the complex permeability and complex permittivity of composite materials show an overall dependence with the volume fraction of the filler.

To obtain the impedance matching condition of the single-layered absorber, the graphical map method has been used [5, 12]. This method assumes that the tan $\delta_\varepsilon = \varepsilon''/\varepsilon'$ is constant in the frequency range, and provides the matching condition for ferrite and ferrite-polymer composite materials. However, the tan $\delta_\varepsilon$ is not constant in this study since the $\varepsilon'$ and $\varepsilon''$ vary with both frequency and filler volume fraction in MnZn ferrite-rubber composites as shown in Fig. 4.

Therefore, in the present paper, we calculated the minimum value of reflection loss with sample thickness and frequency to obtain the matching condition by substituting the measured complex permeability and permittivity of MnZn ferrite-rubber composites into Equations (1) and (2).

Fig. 5 shows the variation of the measured reflection losses of the studied RAM with thickness of 3.0 mm and various MnZn ferrite volume loadings ($v_f$ = 0.10, 0.15, 0.20, 0.27 and 0.37).

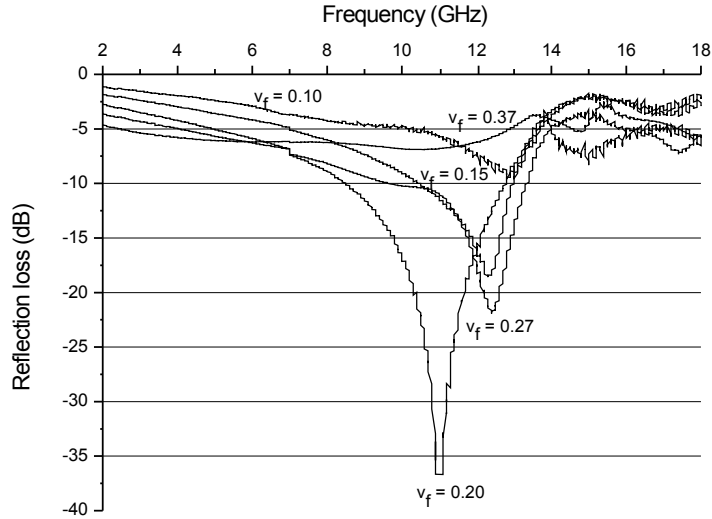

Fig. 5. Effects of MnZn ferrite volume fraction on the reflection loss of the RAM ($t$ = 3.0 mm).

The RAM samples here considered, with different MnZn ferrite contents, have a single-peaked-minimum curve. The reflection loss at the peak is the maximum attenuation of the incident wave and indicates the frequency at which the material offers its optimum wave attenuation properties. The frequency of the minimum reflection loss ($f_m$) is shifted to a lower frequency band with the increase of the volume fraction of MnZn ferrite ($v_f$ = 0.10 → $f_m$ = 13.0 GHz; $v_f$ = 0.15 → $f_m$ = 12.0 GHz; $v_f$ = 0.20 → $f_m$ = 9.0 GHz). As the MnZn ferrite volume fraction increases, the minimum reflection loss decreases from -5 dB for $v_f$ = 0.10 to -35 dB for $v_f$ = 0.20.

According to Feng et al. [13], desirable RAM properties should comply with wider frequency bandwidths of RL < - 10 dB. Therefore, in the design of RAM with better absorption performance, the requirement of wider frequency bandwidths should be observed. Frequency bandwidths can be calculated by subtracting the higher frequency from the lower one at a given RL. We have performed this calculation for 3 arbitrary values of RL as an illustration of our analysis, by referring to the $v_f$ = 0.20 case. The detailed data are shown in Table 1. According to our analysis (c.f. Fig. 5 and Table 1) the MnZn ferrite volume fraction in the RAM is not proportional to the frequency bandwidth. The present RAM, developed from the MnZn ferrite of $v_f$ = 0.20, has the largest frequency bandwidth of 4.5 GHz at RL = -10 dB, whereas those containing MnZn ferrites of $v_f$ = 0.15 and $v_f$ = 027 are 2.5 GHz and 3.0 GHz, respectively.

Table 1 – Frequency bandwidth for the RAM sample with $v_f$ = 0.20.

| RL (dB) | Lower Frequency (GHz) | Higher Frequency (GHz) | Frequency Bandwidth (GHz) |
|---|---|---|---|
| -10 | 8.5 | 13.0 | 4.5 |
| -20 | 10.0 | 12.0 | 2.0 |
| -30 | 10.5 | 11.5 | 1.0 |

In our present study, a trend between volume fraction and reflection loss is not clear. For example, from $v_f$ = 0.10 to 0.20, the amplitude of the RL peak increases with $v_f$, but then decreases for $v_f$ = 0.27 and 0.37.

In order to investigate the attenuation behavior of radar absorbing materials as a function of the thickness, samples with MnZn ferrite of $v_f$ = 0.10 and 0.20 and various thickness values (t = 1.0, 2.0, 3.0, 4.0, 5.0, 6.0, 7.0 and 8.0 mm) were simulated. Their reflection losses were calculated and the results are shown in Fig. 6(a-b). It is clear that the reflection loss of the RAM presents a regular trend at given MnZn ferrite volume fraction in the frequency range of 2 – 12 GHz. Note that the absorption peak of the RAM with MnZn ferrite of $v_f$ = 0.10 (t ≤ 3.0 mm) cannot be perceived in the 2 - 12 GHz range, from the curvilinear trend in Fig. 6(a). In this case, it is observed that the attenuation values increases as the frequency increases with a probable presence of a minimum reflection loss peak in frequencies above 12 GHz. For thickness values above 4,0 mm, the frequency of the minimum reflection loss moves towards the low frequency range with the thickness increase.

When the volume fraction of MnZn ferrite is 0.20, the reflection loss of the radar absorbing material shows the same tendency as that of $v_f$ = 0.10. Note that the amplitude of the main RL peak decreases for smaller thicknesses for $v_f$ = 0.10, in the frequency range analyzed. For $v_f$ = 0.20, this trend seems to reverse for t ≤ 4.0 mm, as can be seen
from a large absorption peak for t = 3.0 mm (Fig. 6(b)).

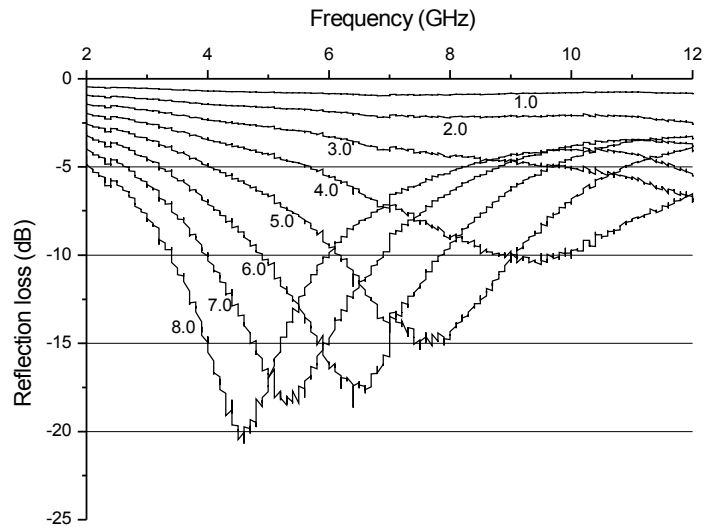

(a)

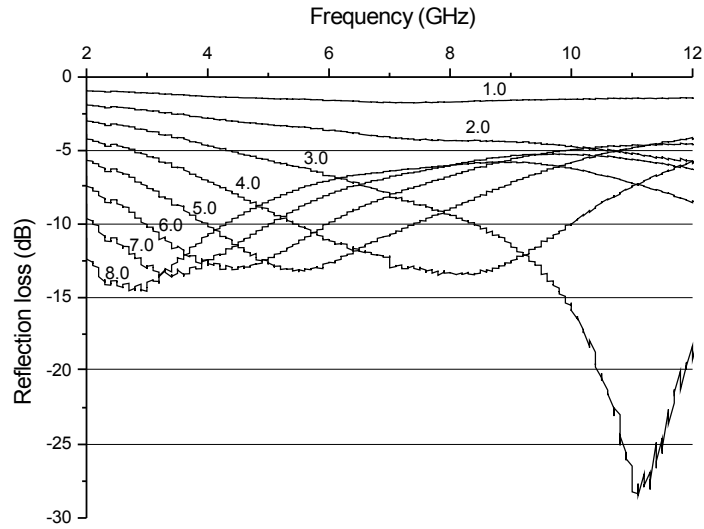

**(b)**

Fig. 6. Calculation of the RAM thickness effects on the microwave absorption properties for: (a) $v_f$ = 0.10 and (b) $v_f$ = 0.20.

In conclusion, our results are in overall agreement with the literature [11 - 13] in the sense that, when estimating the reflection loss of the RAM by the impedance matching degree, in which the former is determined by the combination of the six parameters $\mu'$, $\mu''$, $\varepsilon'$, $\varepsilon''$, $\lambda$ and $t$, it is found that the reflection loss cannot be reduced by merely changing only the $\mu_r$ and $\varepsilon_r$ parameters.

## 4. Conclusion

The results of our study on the electromagnetic and wave attenuation properties of the MnZn ferrite-rubber radar absorbing materials can be summarized as follows.

1) The spectra of the complex magnetic permeability and dielectric permittivity of a MnZn ferrite-silicone rubber radar absorbing material, in the frequency range of 2 - 18 GHz, are provided, in order to establish their quantitative microwave absorption characteristics.

2) We found that a higher MnZn ferrite volume fraction generally favors the increase of the complex relative permeability and relative permittivity of the processed RAM.

3) A prediction of the minimum RL as a function of frequency, sample thickness range and MnZn ferrite volume fraction is given. The results show a regular trend of the RL.

4) With respect to those parameters, such that higher thicknesses imply better absorption properties, with the optimum frequency band shifted systematically towards lower values of frequency. However, for higher $v_f$'s it is inferred that smaller thicknesses might offer better absorption properties for higher frequency bands ($\geq$ 10 GHz).

## 5. Acknowledgements

The authors acknowledge the financial assistance of Financiadora de Estudos e Projetos (FINEP) (Process number 1757-6) and the National Council for Research and Development (CNPq) (Process number 305478/09-5).

## 6. References

[1] L. J. Deng, J. L. Xie, D. F. Liang, B. J. Guo, J. Funct. Mater. 30 (2) (1999) 118.

[2] A. M. Gama, M. C. Rezende, Journal of Aerospace Technology and Management, 2 (1) (2010) 59.

[3] A. N. Yusoff, M. H. Abdullah, S. H. Ahmad, S. F. Jusoh, A. A. Mansor, S. A. A. Hamid, J. Applied Physics 92 (2) (2002) 876.

[4] Y. Naito, K. Suetake, IEEE Trans. MTT 19 (1) (1971) 65.

[5] H. M. Musal, Jr., H. T. Hahn, IEEE Trans. Mag. 25 (5) (1989) 3851.

[6] J. Y. Shin, J. H. Oh, IEEE Trans. Magn. 29 (1993) 3437.

[7] S. S. Kim, S. B. Jo, K. I. Gueon, K. K. Choi, J. M. Kim, K. S. Churn, IEEE Trans. Magn. 27 (1991) 5642.

[8] O. Acher, M. Ledieu, Journal of Magnetism and Magnetic Materials 258-259 (2003) 144-150.

[9] O. Acher, A. L. Adenot, U. S. Patent 6,677,762, (1974).

[10] P. Bartley, S. Begley, S. Materials measurement (2006) 78p. In: <http://www.die.uniroma1.it/personale/frezza/biblioteca/dispense/MisureMateriali.pdf>.

[11] J. Shenhui, D. Ding, J. Quanxing, IEEE, (2003) 590-595.

[12] H. M. Musal, Jr., D. C. Smith, IEEE Trans. Mag. 26 (1990) 1462.

[13] Y. B. Feng, T. Qiu, C. Y. Shen, X. Y. Li, IEEE Trans. Magn. 42 (3) (2006) 363.